\documentclass [12pt]{article}

\begin{document}
\textbf{A hypothesis of earth quake}

{Yeong-Shyeong Tsai}

{Department of Applied Mathematics, National Chung Hsing University,
Taichung,Taiwan}

\textbf{Abstract}

Without a model, it is impossible for a geophysicist to study the
possibility of forecasting earth quakes. We will define a
quantity, the event-degree, in this paper. The quantity plays an
important role in the model of quakes forecasting. In order to
make a simple model, we make a hypothesis of earth quakes. The
hypothesis is: "(i) There are two kinds of earth quakes, one is
the triggered breaking (earth quake), the other is spontaneous
breaking (earth quake). (ii) Most major quakes in continental
plates such as Eurasian Plate, North America Plate, South America
Plate, Africa Plate and Australia Plate are triggered breaking.
(iii) These triggered quakes are triggered by the evolution of
high pressure centers and low pressure centers of the atmosphere
on the plates. (iv) How can the evolution of the high pressure
centers trigger a quake in quantitative sense? It depends on the
event-degree, the extent of the high pressure center, the rate of
evolution of event-degree and the the degree of stored energy for
reaching breaking point."

  The scale of earth is so large that can not be obtained in the
laboratory. Compare to a hurricane, the power of a tornado is very
small though the surface force which is produced by a tornado on
the ground is devastated. compare to the high pressure of
atmosphere, the surface force which is produced by a tornado is
too tiny. Therefore, the surface force which is produced by a high
pressure center of atmosphere might be the trigger of quake.

\textbf{Introduction}

Formosa, Taiwan, is located at the boundary of two plates,
Eurasian plate and Philippine Plate. Therefore, the frequency of
quake is very high. There are more than 4000 quakes in a year.
Only few of them are felt quakes. Compare to the felt quakes, most
of these quakes can be considered as noises in the sense of signal
processing. From the view-point of the weather map, Formosa is
located in the subtropical zone. Roughly speaking, there are only
two seasons: one is dry season and the other is rainy season. From
February to October is rainy season and hence the other period is
dry season. Strictly speaking, the rainy season is from February
to July. In this period, Pacific high pressure center is
comparable to Mongolia high pressure center. These two high
pressure centers confront each other and the boundary of these two
centers passes through the region near by Formosa. The boundary is
called front. During the rainy season, there many fronts which are
built and pushed to southeast and then the front will die out in
several days. Usually, the last front will stay one, two, three or
four weeks. The rainy days happen randomly from the middle of July
to October because they are introduced by typhoons. Though the
history of Taiwan is more than 8000 years, the data collected by
residents is available in the nearest 400 years. People were not
afraid of quake because they were used to the quakes and the
houses were made of bamboos and grasses. The people cared about
the weather because most Taiwans were farmers 100 years ago. When
dry season is too long it becomes draught. When rainy season is
too long there will be no harvest and, even worse, it will flood.
The starting point and the ending point of rainy season is not
predictable. It might come earlier or later. There was not weather
forecasting center at that time, 400 years ago. It seems that the
quakes have nothing concern with the weather. Actually, it might
not be true. From the experiences of Taiwans, a connection between
these two natural phenomena, quakes and weather.

\textbf{A Controversial Issue}

It is unbelievable that they can predicted the starting point and
ending point of the rainy season one day before. Why and how can
they predict these points one day before? We find that they have
got a rule: if there is quake, then this quake is a turning point
of of two seasons, changing rainy season to dry season or changing
dry season to rainy season. The author have tested rule for 50
years. From the view-point of science or, more precisely,
statistics, this is an empirical rule really. Here we should
stress that these quakes are strong felt quakes. This shows the
evidence that there is correlation between weather and quake.
Since we have been confused by this rule, we analyze the data of
weather carefully. We get a rough conclusion: On contrary, we can
use this rule to predict the quake. Of course, this must be
controversial issue. We think the conclusion is reasonable and it
is harmless at least. So far, though there many methods for
predicting quakes, none of them gives a scientific basis or model.
What we mean science is a method or procedures which can be
repeated or, more precisely, verified in any time and at any
place. Since the quake recorders are so sensitive that we must
ignore most of the collected data. It is impractical to predict
such tiny quakes. The first step of our approach is to make a
classification of quakes. Roughly, we classify the quakes two
classes. One class is predictable and the other is unpredictable.
This is not the key point because we think most, if not all, felt
quakes such as the devastated quakes are in the predictable class.

\textbf{A True Story}

 On April 13, 1999, the author gave a talk in press conference.
Most audiences were professors except one or two reporters. The
topics is "the
 empirical rule and the model of quake forecasting." At the end of
 the talk I gave a warning that
 there might be a major quake before the July or
after the September because it was 64 year ago that there was a
major quake in 1935. A reporter said that you must be jerking. All
professors agreed that this was, so to speak, only a hypothesis. I
was assigned the work to find or to collect enough data to verify
the hypothesis. On one hand, it was not available for me to get
the financial support for setting up a large set of global
position systems. On the other hand, it was not easy work to
process the data collected by GPS and the data from weather
forecasting centers. What I could do was to read any related
documents. Usually, the Mongolia high pressure center is not so
strong and active in summer. In summer, the high pressure center
is strong and dominates on the Pacific ocean. Since the water of
ocean will balance itself, it will not disturb the boundary of
plates too much. We found that the Mongolia high pressure center
was strong and active in beginning of the September, 1999. From
our experiences, it was unusual. In the afternoon of September 20,
1999, people were scared why the weather got worse suddenly, the
sky was covered by thick clouds and it started to blow strong wind
(gale). Usually, it must be a shining day in September except a
typhoon near the region. But we did not know what would happen.
Unfortunately, a major quake occurred at 1:47, local time, in the
morning of September 21, 1999. At that moment I was reading the
book "  An Introduction to the theory of Seismology" written by K.
E. Bullen, a text book of my second year course in graduate
school. Since I was exhaustive and I was lying to read the book. I
was almost killed by a falling book case. Fortunately, I was
shocked and stood up. It was painful because one my leg was
injured but not too serious. The distance between my house and the
epicenter of that quake is about 40 kilometers. There were about
2000 people were killed in, the short time, 40 second quake. The
death toll is about 3000 and tens of thousands were injured in
that quake. The Central Weather Bureau announced that the depth of
focus the quake is less than 9 kilometers. The density of
distributed seismic data recorders in Taiwan is the highest one in
the world. An officer of the Central Weather Bureau said very
proudly that the system of data processing was very powerful.
Since most recorders are connected to the computer center of
Weather Central Bureau. The monitoring system of quakes was
computerized completely. The response time of computing system was
less few seconds after the quake. Since the people of Taiwan had
realized that the damage of quakes might happen in any time, the
government spent a lot of money to set up the monitoring system.

\textbf{A Model for Predicting Quake}

Most geophysicist agree with the plate tectonic theory. Therefore,
we will not repeat it in this paper. Though we know that the
energy of quake is supplied from the interior of the earth, we do
not know when the energy will release itself. By analogy, we
construct a simple model or example: A rubber is stretched by
pulling the two ends of the rubber. The forces are large enough
but not too large to break the rubber. We can either increase the
forces slowly or keep the forces constantly. Finally, the rubber
will break. We call it spontaneous breaking (quake). By a finger,
we can disturb or exert a small force perpendicularly at the
middle point of the rubber, then the probability to break the
rubber is very high. If the rubber break, then we call it
triggered breaking ( quake). The situation is quite clear. (I) It
is doubtless that boundary of two plates store the energy for the
breaking (quake). But no one knows when breaking point is reached
and then break. (II) Since there are many forces exert on or
inside the earth such as Newtonian gravity, including celestial
forces, current flow of mass and hence heat inside the earth, the
system of earth is dynamic not static. We can not take them into
account because most of them are unaccessible (III) There are two
kinds of forces: One is body force, like gravity force, the other
is surface force, like pressure force. All objects with mass,
human's cells, a stone, an egg etc., must be exerted by gravity
force, the body force. (IV) From our experiences, the surface
force plays an important role in the triggered breaking. Now, we
try to expose the model for the triggered quake. When we pick up
an egg there are surface forces which exert on the egg. If we
either hold it carelessly or throw it on the ground, then the egg
will be broken. If we put the egg in the water deeply, the
probability of breaking the egg is very low though there are large
surface forces exert on egg. It seems that the trigger force
neither very large nor random happening. Since it takes time to
accumulate the enough energy in the boundary of plates, there is
long period that the boundary of two plates near the breaking
point. Like the rubber in the example, the probability of being
triggered to break is much higher than that of breaking itself.
Therefore, a small scale external force might be a trigger of a
devastated quake if it happens at right time and at right place.

  The scale of earth is so large that can not be obtained in the
laboratory. Compare to a hurricane, the power of a tornado is very
small though the surface force which is produced by a tornado on
the ground is devastated. compare to the high pressure center of
atmosphere, the surface force which is produced by tornado is too
tiny. Therefore, the surface force which is produced by a high
pressure center of atmosphere might be the trigger of quake.

  In speaking of the triggers of the earth quakes, we assume that
the tidal wave will not be the major trigger since the tidal wave
is a balance dynamics, in some sense, and the earth is covered by
water mostly. Likely, when a high pressure atmosphere exerts on
the surface of ocean the water will balance itself automatically,
redistributing the water by flowing.
 When the high pressure
atmosphere force exerts on the continent, the situation is
different. The growth and movement of high (low) pressure of the
atmosphere on the continental plate will trigger a quake if the
energy is stored enough. Of course, if it is necessary, then the
the tidal wave must be considered and this model will be more
complicated and completed. First, in order to make a simple model,
we have made a hypothesis. The force of tidal wave is excluded in
this hypothesis. If it works, then we shall combine effects of
these two forces, the force of tidal waves and the forces of
atmosphere. In order to construct a simple practical model, we
filter out most quakes by classifying them into noises. We hope or
we assume these noise quakes are harmless. This assumption is
reasonable. The reasons are: "(a) The damage of quake is not
dependent on the magnitude of the quake only. Some quakes with
magnitude more than 8 in Richter scale are harmless since the
locations of their focuses are very deep below the ocean area. (b)
Clearly, the deeper is the focus, the less is the probability of
being triggered by the surface force on the ground. (c) Usually,
The devastated quakes are shallow quakes. It is obvious that the
shallow quakes are triggered by the forces of the atmosphere. (d)
The depth of the focus of shallow quake might be less than 10
kilometers. Oil company can drill oil well more than 4 kilometers
in depth. Clearly, the focus of the shallow quake is located at
the region near the ground surface."

  It is possible to watch the sky and find the following condition:
there are two layers of clouds. One layer is higher and other
layer is lower. Each layer consists many pieces of clouds. The
clouds of upper layer are almost stationary while the clouds of
lower layer are moving and changing their shapes. Some times one
cloud of lower layer is almost congruent to one cloud of upper
layer in size and shape and they overlap themselves completely. By
analogy, We construct a model for quake prediction. On a globe, we
model two overlap maps. One is the map of the plate tectonic
theory. There are many plates and their boundaries on this map.
The other is the map of weather. There are many high pressure
centers and their boundaries on the map. Usually, most, if not
all, the low pressure can be consider as the boundaries of two
high pressure. From the view-point of shapes and scales, we have a
correspondence. The correspondence is that the plates are
corresponding to the high pressure (centers) and the boundaries of
the plates are corresponding to the boundaries of high pressure
(centers). The region of this boundary  consists of one, two or
several small scale low pressure centers. Some time, the
boundaries of two high pressure centers are called fronts. We can
not expect the abstract correspondence being so nice that these
two maps, the map of plates and the map high pressure centers,
match perfectly, that is,  each high pressure center is located on
a plate and they are congruent to each other. We assume that the
map of plates is fixed on the globe. The map of weather is
changing in shapes since the weather is changing every moment. It
is possible that some of the high pressure centers are located on
some corresponding plates and they are almost congruent in
geometry. For example, two high pressure centers are located on
two adjacent plates and the geometry shapes of high pressure
centers are almost, or partially, congruent to that of plates
correspondingly. We call this event weather triggering event. In
order to to measure the degree of congruence of the event, a
calculated positive number can be assigned to the congruence.
Since the set of positive number have their natural order. We call
this positive number the degree of the weather trigger event or
the event-degree briefly. Of course, the total force of the high
pressure on the corresponding plate must be a factor of the
event-degree. Therefore, we make the following hypothesis.

\textbf{The Hypothesis}

\textit{"(i) There are two kinds of earth quakes, one is the
triggered breaking (earth quake), the other is spontaneous breaking
(earth quake). (ii) Most major quakes in continental plates such as
Eurasian Plate, North America Plate, South America Plate, Africa
Plate and Australia Plate are triggered breaking. (iii) These
triggered quakes are triggered by the evolution of high pressure
centers and low pressure centers of the atmosphere on the plates.
(iv) How can the evolution of the high pressure centers trigger a
quake in quantitative sense? It depends on the event-degree, the
extent of the high pressure center, the rate of evolution of
event-degree and the the degree of stored energy for reaching
breaking point."}

\textbf{How to Test the Hypothesis}

In signal processing, most works are filtering the noises. In the
hypothesis, we have separated the noise (spontaneous breaking)
from the signal (the triggered breaking). Now, we have a simple
model for testing. Since global position system is available, we
can set up GPS regional widely and collecting data to compute
deformation of plates which are effected by high (low) pressure
atmosphere forces, including the rate of the evolution these high
(low) centers. Both worldwide data of atmosphere pressure and
quake are available. From the records of occurrence time of
quakes, it is possible to test the hypothesis and to modify the
model for forecasting the the earth quakes. Further more, we are
able to find proper way for calculating the event-degree since we
have not had the quantitative results of the event-degree.

\textbf{References}

 [1] Diamond, Jared M. "Taiwan's gift to the world".
Nature, Volume 403, February 2000, pp. 709-710

[2] Korgen Ben J (1995). " {A Voice From the Past: John Lyman and
the Plate Tectonics Story}" . \textit{Oceanography} \textbf{8} (1):
19-20.

[3] Read HH, Watson Janet (1975). \textit{Introduction to Geology}.
Halsted, 13-15

\end{document}